\documentclass[letter]{aa}
\usepackage{txfonts}
\usepackage{natbib,twoopt}
\usepackage{graphicx}
\usepackage{arydshln}
\usepackage{enumitem}
\usepackage{multirow}
\usepackage{multicol}
\usepackage{hyperref}
\usepackage{url}
\usepackage{color}
\bibpunct{(}{)}{;}{a}{}{,} 

\hypersetup{
	pdftitle={CEMP 3D models},
	pdfauthor={A. J. Gallagher},
	pdfsubject={A\&A letter},
	pdfcreator={A. J. Gallagher},
		colorlinks, 
		citecolor=blue, 
		filecolor=blue, 
		linkcolor=blue,
		breaklinks=true, 
		plainpages=false,
		urlcolor=blue 
}

\title{An in-depth spectroscopic examination of molecular bands from 3D hydrodynamical model atmospheres}
\subtitle{II. Carbon-enhanced metal-poor 3D model atmospheres}

\author{A. J. Gallagher\inst{1}\thanks{Observatoire de Paris fellow}
\and
E. Caffau\inst{1}
\and
P. Bonifacio\inst{1}
\and 
H.-G. Ludwig\inst{2,1}
\and
M. Steffen\inst{3,1}
\and
D. Homeier\inst{2}
\and
B. Plez\inst{4}
}

\institute{
GEPI, Observatoire de Paris, PSL Research University, CNRS, Universit\'{e} Paris Diderot, Sorbonne Paris Cit\'{e} Place Jules Janssen, 92190 Meudon, France.\\ 
email: \texttt{andrew.gallagher@obspm.fr}
\and 
Zentrum f{\"u}r Astrononmie der Universit{\"a}t Heidelberg, Landessternwarte, K{\"o}nigstuhl 12, 69117 Heidelberg, Germany.
\and 
Leibniz-Institut f{\"u}r Astrophysik Potsdam, An der Sternwarte 16, 14482 Potsdam, Germany.
\and
Laboratoire Univers et Particules de Montpellier, LUPM, Universit{\'e} de Montpellier, CNRS, 34095 Montpellier cedex 5, France
}

\date{Received ... / Accepted ...}

\authorrunning{A. J. Gallagher et al.}
\titlerunning{3D CEMP model atmospheres}

\newcommand{\cobold}{CO$^{5}$BOLD}
\newcommand{\atd}{$\langle{\rm 3D}\rangle$}
\newcommand{\odx}{{\tt LHD}}
\newcommand{\teff}{T_{\rm eff}}
\newcommand{\logg}{\log{g}}

\newcommand{\feh}{{\rm [Fe/H]}}
\newcommand{\logtr}{\log{\tau_{\rm ROSS}}}

\newcommand{\actd}{A({\rm C})_{\rm 3D}}

\newcommand{\ac}{A({\rm C})}
\newcommand{\an}{A({\rm N})}
\newcommand{\oa}{A({\rm O})}

\newcommand{\ma}{{\tt\bfseries A}}
\newcommand{\mb}{{\tt\bfseries B}}
\newcommand{\mc}{{\tt\bfseries C}}
\newcommand{\sab}{{\tt\bfseries SAB}}
\newcommand{\sac}{{\tt\bfseries SAC}}
\newcommand{\smb}{{\tt\bfseries SB}}
\newcommand{\smc}{{\tt\bfseries SC}}
\newcommand{\gband}{{\it G}-band}

\defcitealias{Gallagher2016a}{Paper I}
\newcommand{\pp}{\citetalias{Gallagher2016a}}
\abstract
{Tighter constraints on metal-poor stars we observe are needed to better understand the chemical processes of the early Universe. Computing a stellar spectrum in 3D allows one to model complex stellar behaviours, which cannot be replicated in 1D.}
{
We examine the effect that the intrinsic CNO abundances have on a 3D model structure and the resulting 3D spectrum synthesis.
}
{
Model atmospheres were computed in 3D for three distinct CNO chemical compositions using the \cobold\ model atmosphere code, and their internal structures were examined. Synthetic spectra were computed from these models using Linfor3D and they were compared. New 3D abundance corrections for the \gband\ and a selection of UV OH lines were also computed.
}
{
The varying CNO abundances change the metal content of the 3D models. This had an effect on the model structure and the resulting synthesis. However, it was found that the C/O ratio had a larger effect than the overall metal content of a model.
}
{
Our results suggest that varying the C/O ratio has a substantial impact on the internal structure of the 3D model, even in the hot turn-off star models explored here. This suggests that bespoke 3D models, for specific CNO abundances should be sought. Such effects are not seen in 1D at these temperature regimes.
}

\begin{document}

\maketitle

\section{Introduction}
\label{sec:intro}

Carbon-enhanced metal-poor (CEMP) stars are a class of metal-poor star that exhibit over-abundances in carbon, relative to their iron abundance; ${\rm [C/Fe]}>+1.0$ \citep{Beers2005}. To date, CEMP stars represent roughly $1/5^{\rm th}$ of the known population of metal-poor stars \citep{Lucatello2006}. They can be further categorised according to their neutron-capture element abundances, giving rise to four sub-classes: CEMP-no, CEMP-r/s, CEMP-r and CEMP-s. Formal definitions of these sub-classes are given in \citet{Beers2005}. \citet{Spite2013} found that when the absolute carbon abundances, $A({\rm C})$, of CEMP stars are presented as a function of $\feh$, two distinct bands are found; the high and low carbon bands. This work was subsequently extended in \citet{Bonifacio2015}. The high carbon band contains the most iron-rich CEMP stars, and is populated by all four sub-classes of CEMP star. Evidence suggests that most, if not all these stars attained their high $\ac$ through possible binary interaction; typically, these stars show radial velocity variations indicative of a faint companion \citep{Lucatello2005}. None of the stars on the low carbon band demonstrate an overabundance in the heavy elements, although the available upper limits on [Ba/Fe] are not precise enough to allow a firm classification as CEMP-no stars. It would appear that these stars formed from gas clouds that were previously enriched with carbon. However, the process(es) by which these gas clouds procured their enhanced $\ac$ is still highly debated \citep[see][and references therein]{Salvadori2015}. Therefore, tighter constraints from observations are highly sought so that the chemical processes in the early Universe can be better understood. This in turn requires the use of more sophisticated modelling of the stellar atmosphere to compute more realistic synthetic spectra. 

\begin{table*}[!ht]
\caption{Model and syntheses abundances used in the present investigation. All models were computed using identical stellar parameters $\teff/\logg/\feh=6250\,{\rm K}/4.0/-3.0$.}
\begin{center}
\begin{tabular}{l c c c c c c l c c c c r}
\hline\hline
Model & \multicolumn{4}{c}{Model abundances} & Model & & Synthesis & \multicolumn{4}{c}{Syntheses abundances} & Comment \\
\cline{2-5} \cline{9-12}
ID & $\ac$ & $\an$ & $\oa$ & C/O & $\Delta{\rm t}$ (s) && ID & $\ac$ & $\an$ & $\oa$ & C/O \\
\hline
\ma\ & $5.39$ & $4.78$ & $6.06$ & $0.21$ & $50600$ && \sab\ & $7.39$ & $6.78$ & $6.06$ & $21.4$ & Inconsistent \\
      &        &        &        & &  && \sac\ & $7.39$ & $6.78$ & $7.66$ & $0.54$ & Inconsistent \\
\hdashline
\mb\ & $7.39$ & $6.78$ & $6.06$ & $21.4$ & $45200$ && \smb\ &  $7.39$ & $6.78$ & $6.06$ & $21.4$ & Consistent \\
\mc\ & $7.39$ & $6.78$ & $7.66$ & $0.54$ & $71600$ && \smc\ &  $7.39$ & $6.78$ & $7.66$ & $0.54$ & Consistent \\
\hline
\end{tabular}
\label{tab:models}
\tablefoot{Abundances of syntheses \sab\ and \sac\ computed using model \ma\ were artificially enhanced to match the intrinsic abundances of models \mb\ and \mc, making them inconsistent with the intrinsic CNO abundances of model \ma. This is how 3D corrections were computed in \citetalias{Gallagher2016a}. ``Model $\Delta{\rm t}$'' refers to difference in time (in seconds) between the first and last snapshot selected in the model series.}
\end{center}
\end{table*}

In the first paper in this series, \citep[henceforth \citetalias{Gallagher2016a}]{Gallagher2016a}, we recently published a grid of 3D $\ac$ corrections for several dwarf star models. They were computed by fitting 1D synthetic \gband\ profiles to 3D \gband\ profiles, essentially treating the 3D synthetic profiles as observed data of known $\ac$. These syntheses were computed using the temperature structures of 3D model atmospheres that have normal metal-poor chemical compositions, and the CNO abundances were then enhanced for the spectrum synthesis. This presented clear inconsistencies between the intrinsic chemical composition of the model and the  chemical composition of the synthetic spectra.

It is common practice to assume that the iron abundance of a star, $\feh$, is equal to its metal content, so that $\log{Z/{\rm Z}_\odot}\approx\feh$. A simple calculation shows that while an $\feh=-3.0$ metal-poor star with $[\alpha/{\rm Fe}]=+0.4$ has a metal content by mass fraction of $Z\approx2.75\times10^{-5}$ (or $\log{Z/{\rm Z}_\odot}\approx -2.71$), a CEMP star with $\Delta A({\rm CNO})=+2$\,dex, with the same $\feh$, has a $Z\approx1.66\times10^{-3}$ ($\log{Z/{\rm Z}_\odot}\approx -0.93$); almost two orders of magnitude larger. It was already demonstrated in \citetalias{Gallagher2016a} that the 3D temperature gradient (as a function of $\logtr$) of a metal-poor model is steeper than a more metal-rich model, and that molecular formation is heavily influenced ultimately by the temperature structure of the model. Molecular opacities in turn influence the temperature structure of the model. Therefore, it could be argued that the 3D metal-poor models utilised in \citetalias{Gallagher2016a} were not appropriate to compute \gband\ spectra of CEMP stars. We present a significant test that examines whether intrinsic 3D CEMP models are necessary for modelling molecular spectral features in stars with peculiar CNO abundances. We also assess this supposition by computing new 3D corrections for the synthetic \gband\ using these new models and compare them to corrections computed from spectra synthesised using a metal-poor model.  It has been shown that CNO abundances only matter in 1D at lower temperatures \citep{Plez2005, Gustafsson2008, Masseron2008}.

\section{Model atmospheres and spectrum synthesis}
\label{sec:modelsandsynth}

The work presented here required the computation of 3D model atmospheres. They were computed using \cobold\ \citep{Freytag2012} using the box-in-a-star mode. All models were computed with identical stellar parameters $\teff/\logg/\feh=6250\,{\rm K}/4.0/-3.0$, but the model chemical composition was varied. Our investigation also called for the computation of 3D synthetic spectra. As the features of interest were molecular bands, we computed them with Linfor3D\footnote{\href{http://www.aip.de/Members/msteffen/linfor3d}{http://www.aip.de/Members/msteffen/linfor3d}}, which was recently modified to compute large spectral regions \citep{Gallagher2016b}.  

For the interested reader, Appendix~\ref{apx:1D} presents a brief investigation under 1D, analogous to the one detailed here.

\subsection{\cobold\ model computations}
\label{sec:models}

The \cobold\ 3D model atmospheres consist of a series of 20 computational boxes, referred to as snapshots. The snapshots were selected from a larger temporal sequence of computational boxes that have reached full dynamical and thermal relaxation. The snapshots selected are spaced far enough apart in time ($\sim58$ characteristic timescales\footnote{$t_{\rm c} = H_{\rm P} / c$, where $H_{\rm P}$ and $c$ are the pressure scale height at $\logtr=1$ and sound speed, respectively.}) to be considered statistically independent. The spatial resolution and geometrical size of all atmospheres were $140\times140\times150$ grid points representing a $26.1\times26.1\times12.8\,{\rm Mm}$ region of the stellar atmosphere. Opacities were based on OSMARCS model opacities (these models are used in Appendix~\ref{apx:1D}), which were binned into $14$ opacity groups \citep{Nordlund1982,Ludwig1994}. The effect of scattering was treated according to the \citet{Hayek2010} approximation as explored by \citet{Collet2011}, i.e. scattering opacities in the deeper layers of the model are treated as true absorption and ignored in the outer regions of the model atmosphere \citep[see][for a full description]{Ludwig2013}. As all models used are metal-poor, the alpha elements were enhanced so that $[\alpha/{\rm Fe}]=+0.4$. A summary of the models computed for this investigation, and their intrinsic CNO abundances are presented in Table~\ref{tab:models}. The nomenclature given in this table is used throughout this work for clarity. The opacity sources for each model include the effects associated with changing $A({\rm CNO})$ (e.g. CH, OH, C, etc.). We did not account for changes in the thermodynamic properties of the plasma due to the enhancement of the CNO abundances, and used the same equation-of-state table for all three models. While the change of the relative number of metallic atoms is indeed substantial (increasing from $2\times 10^{-6}$ to $7\times 10^{-5}$), the overall number fraction remains minute. As such, we do not expect a noticeable impact on molecular weight, specific heat, etc. of the plasma, which are mostly controlled by hydrogen and helium. 
\begin{figure*}[!ht]
\begin{center}
\includegraphics[width=\linewidth]{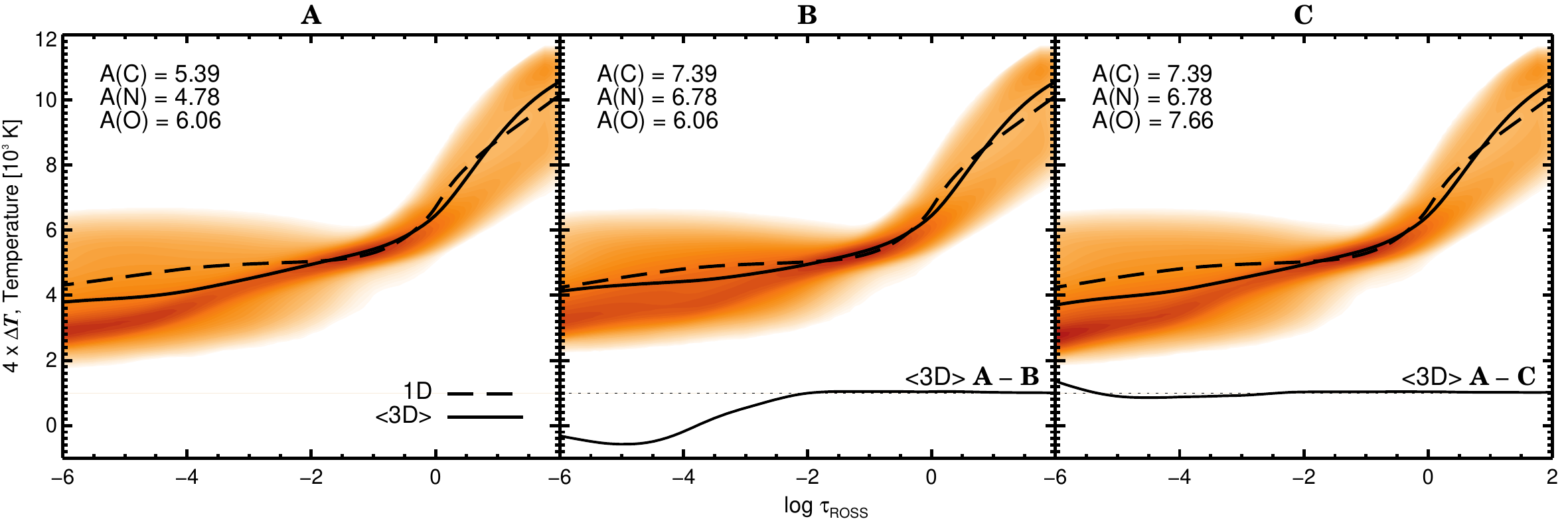}
\caption{Temperature structures of the three model atmospheres analysed in this investigation. The intrinsic CNO abundances are given in each panel. Details about each model and the synthesis produced with them is given in Table~\ref{tab:models}. Panels \mb\ and \mc\ also present the differences between the \atd\ temperature structures of model \ma\ with models \mb\ and \mc\ respectively. This has been scaled up four times relative to the y-axis to better show the temperature deviations so that every tick represents a $\Delta T = 50$\,K.}
\label{fig:tstruct}
\end{center}
\end{figure*}

Each 3D model atmosphere has two counterpart model atmospheres, computed for the same stellar parameters and model chemical composition. The first is an external 1D model atmosphere computed using the LHD model atmosphere code \citep{Caffau2007}. LHD model atmospheres use the same micro-physics approximations (equation-of-state, opacities, etc.) adopted by the 3D models. The second is an averaged, \atd, model atmosphere computed for each 3D snapshot by spatially averaging the thermal structure of the computational box over surfaces of equal Rosseland optical depth. 

\subsection{Spectrum synthesis}
\label{sec:synthesis}

Synthetic spectra for the CH \gband\ ($4140-4400$\,\AA) and the CN BX-band ($3870-3890$\,\AA) features, as well as 14 UV OH transitions ($3122-3128$\,\AA), which we refer to as the OH-band, were computed for this analysis. The line list used to synthesise the \gband\ is described in \citetalias{Gallagher2016a}. The line list used for the CN-band was identical to that detailed in \citet{Gallagher2016b}. The line list of the OH-band was adopted from part of the OH line list used by \citet{Spite2017}.

The chemical compositions input into Linfor3D for the syntheses of the two CEMP model atmospheres, \mb\ and \mc, were fixed to that of the intrinsic chemical structures of the models. As such, a single \gband, CN-band and OH-band spectrum was computed for each of these models: \smb\ and \smc. The chemical abundances used to compute syntheses from model \ma\ were enhanced to match the chemical compositions of models \mb\ and \mc, resulting in the synthesis of two \gband, CN-band and OH-band spectra: \sab\ and \sac, see Table~\ref{tab:models}.

Every 3D synthetic spectrum had a counterpart spectrum computed using the 1D \odx\ and \atd\ atmospheres. The spectra from these two models are computed along with the 3D spectra in Linfor3D to help distinguish the effects the stellar granulation (3D to \atd) and the lower average temperatures (\atd\ to 1D) in the outermost regions have on the resultant spectra.

\section{Results}
\label{sec:results}

Figure~\ref{fig:tstruct} depicts the temperature structures of models \ma, \mb\ and \mc\ and presents how the chemical composition affects the thermal structure of the 3D models, and their counterpart 1D LHD and \atd\ model atmospheres. For clarity, we also present the \atd\ temperature differences between model \ma\ and models \mb\ and \mc\ in the centre and right panels of the figure. The temperature differences of the 1D models are not presented as they are almost identical. It is clear that changes to the chemical composition of a model do impact the 3D temperature stratification. The C/O\footnote{${\rm X/Y}=N({\rm X})/N({\rm Y})=10^{\left[ A({\rm X})_*-A({\rm Y})_* \right]}$. The reader may prefer to consider the $\ac-\oa$ ratio instead.} ratio of models \ma\ and \mb\ differ the most. If they are compared, the temperature fluctuations become smaller in \mb, and the average temperature, \atd, increases by upto $400$\,K in the outermost regions of the atmospheres. However, when models \ma\ and \mc\ are compared, the mean temperature and the temperature fluctuations are very well reproduced. This suggests that the C/O ratio is more important to the model structure than the overall metal content, $Z$. This indicates that when the C/O ratio is high, the blanketing effect of carbon-bearing molecules other than CO affect the temperature structure of the 3D model, which in turn would also affect the absorbing strength of the CH- and CN-bands.

\begin{figure*}[!ht]
\begin{center}
\includegraphics[width=\linewidth]{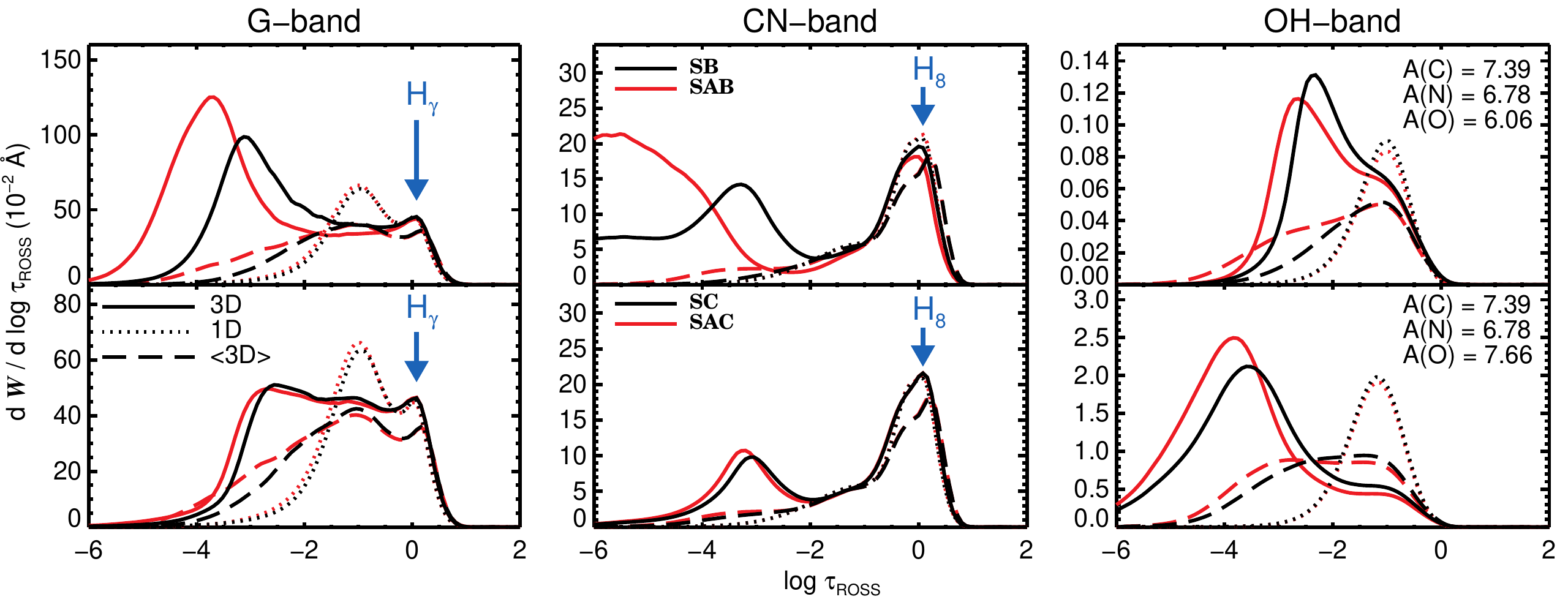}
\caption{3D, \atd, and 1D contribution functions of the G-, CN- and OH-bands. {Top:} comparison between \smb\ and \sab. {Bottom:} comparison between \smc\ and \sac. Indicators have been marked in the \gband\ and CN-band to show where the Balmer features are forming in these bands.}
\label{fig:cfs}
\end{center}
\end{figure*}

Figure~\ref{fig:cfs} depicts the equivalent width contribution functions of the \gband, CN-band and OH-band. It is derived by integrating the line-depth contribution function \citep{Magain1986} over all wavelength points sampled during the spectrum synthesis. An equivalent width contribution function is useful as its integral reproduces the equivalent width of the feature, allowing us to compare the contribution of the various atmospheric layers to the total absorption that makes up a spectral feature. The black lines represent the contribution functions of the spectra \smb\ (top) and \smc\ (bottom), while the red lines are those from \sab\ (top) and \sac\ (bottom). The formations of the \smb\ and \sab\ (i.e ${\rm C/O}=21.4$) CN- and {\it G}-bands are drastically different, while the formation of \smc\ and \sac\ (i.e. ${\rm C/O}=0.54$) is fairly comparable. As was explained in \citetalias{Gallagher2016a}, the oxygen- and carbon-bearing molecules -- more specifically CH and OH -- have an anti-correlated interrelationship. The difference in behaviour between \smc\ and \sac\ in the OH-band is not as significant as is seen in \smb\ and \sab\ for the CH \gband. 
The scenario presented here only alters the C/O ratio by changing $\oa$ while $\ac$ is fixed. Appendix A in \citetalias{Gallagher2016a}  demonstrated that when $\oa$ is fixed and the carbon is allowed to vary, the formation of OH behaves very much like CH, although in an anti-correlated manner.

\begin{table}[!t]
\caption{3D abundance corrections, $\Delta_{\rm 3D} = \actd - A({\rm C})_{\rm 1D, LHD}$, for the \gband\ and the OH-band from the models examined in this work.}
\begin{center}
\begin{tabular}{l c c r}
\hline\hline
Model & Synthesis & \multicolumn{2}{c}{$\Delta_{\rm 3D}$} \\
\cline{3-4}
ID    & ID        &      \gband\      &    OH-band       \\
\hline
\ma\  &   \sab\   &       $-0.76$     &     $-0.40$      \\
      &   \sac\   &       $-0.31$     &     $-0.88$      \\
\mb\  &   \smb\   &       $-0.50$     &     $-0.35$      \\
\mc\  &   \smc\   &       $-0.32$     &     $-0.73$      \\
\hline
\end{tabular}
\label{tab:corrections}
\end{center}
\end{table}

Table~\ref{tab:corrections} provides 3D corrections, $\Delta_{\rm 3D}=\actd-A({\rm C})_{\rm 1D,LHD}$, of the \gband\ and OH-band. Based on the behaviours of \smc\ and \sac\ for the \gband\ in Fig.~\ref{fig:cfs}, and the corrections of \smb\ and \sab\ in Table~\ref{tab:corrections}, the 3D corrections to the \gband\ presented in \citetalias{Gallagher2016a} are still valid, as the C/O ratio used for those syntheses ($0.21$) is comparable to \sac\ and \smc\ ($0.54$). When the C/O ratio is large (case \smb\ and \sab), the 3D corrections are no longer similar and the large correction provided by \sab\ is reduced by $0.26$\,dex by \smb. This also supports the previous work done on the C/O ratio in \pp. The corrections provided by the OH-band further demonstrate the anti-correlated behaviour with CH. When the C/O ratio is small, the excess oxygen not used to form CO, forms more OH. The opposite is seen for the \gband. It was decided that it would be inappropriate to include $\an$ corrections from a comparative analysis of the \gband\ and CN-band because we have not investigated the additional impact $\an$ will have on this band or the model atmospheres, and also because of the extreme behaviour seen in the CN-band formation; the CN-band forms outside the confines of the computational box in syntheses \smb\ and \sab. 

\section{Discussion and conclusions}
\label{sec:conclusions}

We have confirmed that a large, non-solar-like C/O ratio affects the structure of a 3D model (models \ma\ and \mb, Fig~\ref{fig:tstruct}). We have also shown that this impacts the synthetic spectra computed from them (syntheses \smb\ and \sab, Fig~\ref{fig:cfs}). Due to the cooler average temperature of model \ma, relative to \mb, higher number densities of CH and CN are permitted to form in shallower regions when \sab\ is synthesised assuming LTE. The higher temperatures of model \mb\ reduces overall line strength of \smb\  and partially restricts molecule formation in the outer regions, pushing formation inward (see the G- and CN-bands for syntheses \smb\ and \sab, Fig~\ref{fig:cfs}). In this regime, the 3D corrections from \sab\ are overestimated and should be replaced by the consistent corrections from \smb. Figure~\ref{fig:cfs} also demonstrates that, in contrast to 3D, inconsistencies in the chemical composition of the 1D LHD models do not impact the synthetic 1D spectra, as they do not probe the temperature ranges where these models form in large quantities. This means that errors do not cancel in the 3D-1D differential approach because the inconsistencies in the intrinsic chemical composition matter for 3D, but not for 1D.

We have demonstrated in a limited fashion that the overall effect the CNO abundance has on the 3D model structure is small if the C/O ratio remains fairly consistent between the models and syntheses at this temperature regime, as presented by models \ma\ and \mc\ and by syntheses \smc\ and \sac. In this scenario, model \mc\ has a C/O ratio that is approximately twice as large as is found in model \ma, as $\ac$ in model \mc\ is $+2.0$\,dex larger than in model \ma. Nevertheless, synthesis \sac\ appropriately replicates synthesis \smc\ for both carbon molecular bands. This is supported by the consistent 3D corrections found in \smc\ and \sac\ for the \gband, given in Table~\ref{tab:corrections}. At this point in time, it is not possible to empirically determine the C/O value at which a new model must be computed as the syntheses from a model like \ma\ will become unreliable. From the models analysed here, it would seem that the impact that $\ac$ has on the thermal structure is minimal as long as ${\rm C/O}<1$, but it becomes significant when ${\rm C/O}>1$.

We have explored the CN-band in detail. As was foreshadowed by \citet{Bonifacio2013} and \citet{Gallagher2016b}, the abundances determined from the standard 3D model are unreliable when the ${\rm C/O}>1.0$, as a significant fraction of its formation occurs outside the confines of the computational box. We confirm this behaviour in synthesis \sab, and show that it is greatly improved by synthesis \smb\ but not fully resolved. In both instances, ${\rm C/O}=21.4$. Interestingly, we can see that this is a purely 3D effect, due to the stellar granulation, as the \atd\ and 1D syntheses do not trace the behaviour of the 3D contribution functions lower than $\logtr<-2$, and only a very small amount of formation in the region $-5\leq\logtr\leq-3$ can be attributed to the lower average temperature of model \ma\ relative to \mb\ (as shown by the red and black dashed lines in Fig.~\ref{fig:cfs}, top middle panel). However, inspection of the granulation patterns of all three models do not reveal any obvious differences in granulation size or behaviour. It has been established that the CN-band is sensitive to departures from thermodynamic equilibrium (NLTE), and it is also known that part of the CN-band forms in the chromosphere of the Sun \citep{Mount1975}. Studies by \citet{Peterson1997} and \citet{Takeda2011} both confirm that metal-poor dwarf stars have a chromosphere. Further exploration of the CN-band behaviour in 3D is warranted. The additional effect $\an$ has on the formation of the CN-band should also be considered.

This work presents a first step into a new area of research in 3D stellar atmosphere modelling. Based on spectroscopic analyses done in 1D for decades, it was generally understood that the metal abundance pattern of a model (i.e. anything beyond helium) had little-to-no effect on the metal-poor model as the total mass fraction was imperceptibly changed. Under the 3D caveat, we have found that this is also held, but only when the C/O ratio is roughly consistent between model and synthesis. When this changes, the model structure changes enough to affect the resulting spectra computed from them. If this is not taken into account, it can lead to line formation outside the computational box, over- or under-estimations in the strength of the feature, and inappropriate 3D corrections. We intend to extend this work and provide a detailed analysis for a larger number of 3D CEMP models.

\begin{acknowledgements}
This project is funded by FONDATION MERAC and the matching fund granted by the Scientific Council of Observatoire de Paris. We acknowledge support from the Programme National de Cosmologie et Galaxies (PNCG) and Programme  National de Physique Stellaire (PNPS) of the Institut National de Sciences de l'Univers of CNRS. This work was supported by Sonderforschungsbereich SFB 881 "The Milky Way System" (subproject A4) of the German Research Foundation (DFG).
\end{acknowledgements}

\bibliographystyle{aa}
\bibliography{bigbib}

\begin{thebibliography}{25}
\expandafter\ifx\csname natexlab\endcsname\relax\def\natexlab#1{#1}\fi

\bibitem[{{Beers} \& {Christlieb}(2005)}]{Beers2005}
{Beers}, T.~C. \& {Christlieb}, N. 2005, \araa, 43, 531

\bibitem[{{Bonifacio} {et~al.}(2013){Bonifacio}, {Caffau}, {Ludwig}, {Spite},
  {Plez}, {Steffen}, \& {Spite}}]{Bonifacio2013}
{Bonifacio}, P., {Caffau}, E., {Ludwig}, H.-G., {et~al.} 2013, Memorie della
  Societa Astronomica Italiana Supplementi, 24, 138

\bibitem[{{Bonifacio} {et~al.}(2015){Bonifacio}, {Caffau}, {Spite}, {Limongi},
  {Chieffi}, {Klessen}, {Fran{\c c}ois}, {Molaro}, {Ludwig}, {Zaggia}, {Spite},
  {Plez}, {Cayrel}, {Christlieb}, {Clark}, {Glover}, {Hammer}, {Koch},
  {Monaco}, {Sbordone}, \& {Steffen}}]{Bonifacio2015}
{Bonifacio}, P., {Caffau}, E., {Spite}, M., {et~al.} 2015, \aap, 579, A28

\bibitem[{{Caffau} \& {Ludwig}(2007)}]{Caffau2007}
{Caffau}, E. \& {Ludwig}, H.-G. 2007, \aap, 467, L11

\bibitem[{{Collet} {et~al.}(2011){Collet}, {Hayek}, {Asplund}, {Nordlund},
  {Trampedach}, \& {Gudiksen}}]{Collet2011}
{Collet}, R., {Hayek}, W., {Asplund}, M., {et~al.} 2011, \aap, 528, A32

\bibitem[{{Freytag} {et~al.}(2012){Freytag}, {Steffen}, {Ludwig},
  {Wedemeyer-B{\"o}hm}, {Schaffenberger}, \& {Steiner}}]{Freytag2012}
{Freytag}, B., {Steffen}, M., {Ludwig}, H.-G., {et~al.} 2012, Journal of
  Computational Physics, 231, 919

\bibitem[{{Gallagher} {et~al.}(2016{\natexlab{a}}){Gallagher}, {Caffau},
  {Bonifacio}, {Ludwig}, {Steffen}, \& {Spite}}]{Gallagher2016a}
{Gallagher}, A.~J., {Caffau}, E., {Bonifacio}, P., {et~al.} 2016{\natexlab{a}},
  \aap, 593, A48

\bibitem[{{Gallagher} {et~al.}(2016{\natexlab{b}}){Gallagher}, {Steffen},
  {Caffau}, {Bonifacio}, {Ludwig}, \& {Freytag}}]{Gallagher2016b}
{Gallagher}, A.~J., {Steffen}, M., {Caffau}, E., {et~al.} 2016{\natexlab{b}},
  ArXiv e-prints [\eprint[arXiv]{1610.04427}]

\bibitem[{{Gustafsson} {et~al.}(2008){Gustafsson}, {Edvardsson}, {Eriksson},
  {J{\o}rgensen}, {Nordlund}, \& {Plez}}]{Gustafsson2008}
{Gustafsson}, B., {Edvardsson}, B., {Eriksson}, K., {et~al.} 2008, \aap, 486,
  951

\bibitem[{{Hayek} {et~al.}(2010){Hayek}, {Asplund}, {Carlsson}, {Trampedach},
  {Collet}, {Gudiksen}, {Hansteen}, \& {Leenaarts}}]{Hayek2010}
{Hayek}, W., {Asplund}, M., {Carlsson}, M., {et~al.} 2010, \aap, 517, A49

\bibitem[{{Kurucz}(2005)}]{Kurucz2005}
{Kurucz}, R.~L. 2005, Memorie della Societa Astronomica Italiana Supplementi,
  8, 14

\bibitem[{{Lucatello} {et~al.}(2006){Lucatello}, {Beers}, {Christlieb},
  {Barklem}, {Rossi}, {Marsteller}, {Sivarani}, \& {Lee}}]{Lucatello2006}
{Lucatello}, S., {Beers}, T.~C., {Christlieb}, N., {et~al.} 2006, \apjl, 652,
  L37

\bibitem[{{Lucatello} {et~al.}(2005){Lucatello}, {Tsangarides}, {Beers},
  {Carretta}, {Gratton}, \& {Ryan}}]{Lucatello2005}
{Lucatello}, S., {Tsangarides}, S., {Beers}, T.~C., {et~al.} 2005, \apj, 625,
  825

\bibitem[{{Ludwig} {et~al.}(1994){Ludwig}, {Jordan}, \& {Steffen}}]{Ludwig1994}
{Ludwig}, H.-G., {Jordan}, S., \& {Steffen}, M. 1994, \aap, 284, 105

\bibitem[{{Ludwig} \& {Steffen}(2013)}]{Ludwig2013}
{Ludwig}, H.-G. \& {Steffen}, M. 2013, Memorie della Societa Astronomica
  Italiana Supplementi, 24, 53

\bibitem[{{Magain}(1986)}]{Magain1986}
{Magain}, P. 1986, \aap, 163, 135

\bibitem[{{Masseron}(2008)}]{Masseron2008}
{Masseron}, T. 2008, in American Institute of Physics Conference Series, Vol.
  990, First Stars III, ed. B.~W. {O'Shea} \& A.~{Heger}, 178--180

\bibitem[{{Mount} \& {Linsky}(1975)}]{Mount1975}
{Mount}, G.~H. \& {Linsky}, J.~L. 1975, \apjl, 202, L51

\bibitem[{{Nordlund}(1982)}]{Nordlund1982}
{Nordlund}, A. 1982, \aap, 107, 1

\bibitem[{{Peterson} \& {Schrijver}(1997)}]{Peterson1997}
{Peterson}, R.~C. \& {Schrijver}, C.~J. 1997, \apjl, 480, L47

\bibitem[{{Plez} \& {Cohen}(2005)}]{Plez2005}
{Plez}, B. \& {Cohen}, J.~G. 2005, \aap, 434, 1117

\bibitem[{{Salvadori} {et~al.}(2015){Salvadori}, {Sk{\'u}lad{\'o}ttir}, \&
  {Tolstoy}}]{Salvadori2015}
{Salvadori}, S., {Sk{\'u}lad{\'o}ttir}, {\'A}., \& {Tolstoy}, E. 2015, \mnras,
  454, 1320

\bibitem[{{Spite} {et~al.}(2013){Spite}, {Caffau}, {Bonifacio}, {Spite},
  {Ludwig}, {Plez}, \& {Christlieb}}]{Spite2013}
{Spite}, M., {Caffau}, E., {Bonifacio}, P., {et~al.} 2013, \aap, 552, A107

\bibitem[{{Spite} {et~al.}(2017){Spite}, {Peterson}, {Gallagher}, {Barbuy}, \&
  {Spite}}]{Spite2017}
{Spite}, M., {Peterson}, R.~C., {Gallagher}, A.~J., {Barbuy}, B., \& {Spite},
  F. 2017, ArXiv e-prints [\eprint[arXiv]{1701.04608}]

\bibitem[{{Takeda} \& {Takada-Hidai}(2011)}]{Takeda2011}
{Takeda}, Y. \& {Takada-Hidai}, M. 2011, \pasj, 63, 547

\end{thebibliography}

\begin{appendix}

\section{CEMP model atmospheres in 1D}
\label{apx:1D}

We present the results and conclusions from our 1D investigation which was done to confirm that model chemical abundances have little-to-no-effect on the resulting spectrum synthesis for the stellar parameters $\teff/\logg/\feh=6250\,{\rm K}/4.0/-3.0$. We utilised three 1D OSMARCS \citep{Gustafsson2008} and three 1D ATLAS12 \citep{Kurucz2005} model atmospheres. These models have the same three distinct chemical abundances as models \ma, \mb\ and \mc, given in Table~\ref{tab:models}. Syntheses were computed in Linfor3D with these models using the same procedures described in Sect.~\ref{sec:synthesis}. As such, we refer to them using the same nomenclature established above.

\begin{figure}[!h]
\begin{center}
\includegraphics[width=0.95\linewidth]{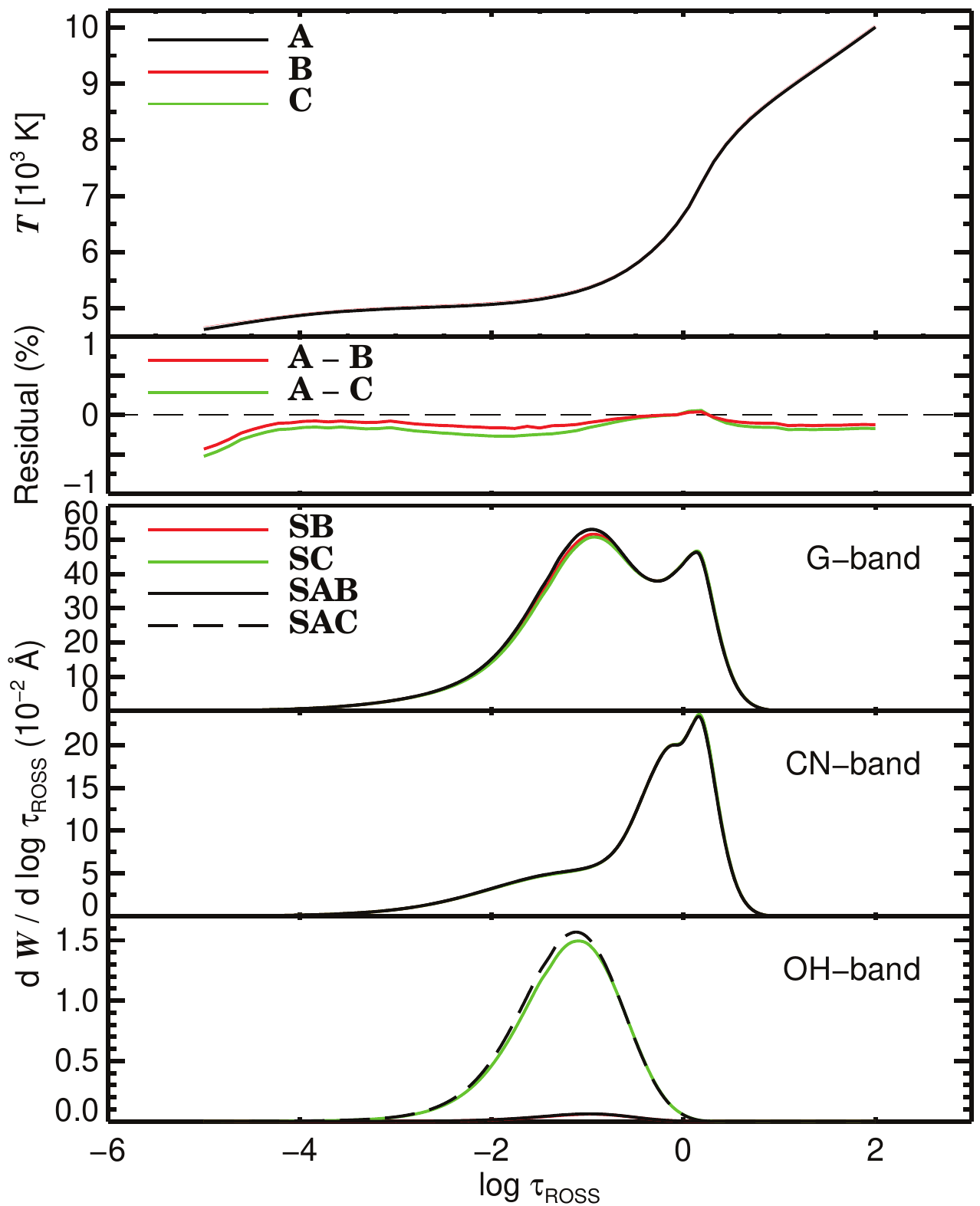}
\end{center}
\caption{1D OSMARCS temperature structures for the model abundances given in Table~\ref{tab:models} and the residuals (top two panels). Resulting contribution functions for the CH- CN- and OH-bands are given in the bottom three panels. The contribution functions for the CN-band are indistinguishable.}
\label{fig:MARCS}
\end{figure}

The temperature structures of the 1D OSMARCS models are presented in Fig.~\ref{fig:MARCS}. For better resolution, the relative temperature difference between model \ma\ with models \mb\ and \mc\ is given as a percentage in the second panel. As depicted, the intrinsic chemical structures have a very limited influence on the temperature structure. In fact, it is shown in the second panel that the differences in temperature between \ma\ and the two CEMP models, \mb\ and \mc, change the temperature by less than $0.5\%$ towards the outermost layers of the model. Between the regions $-4\leq\logtr\leq0$ where lines and continua form in 1D, the changes in the temperatures structures are even smaller.

To test the effect that these minute changes in temperature have on the resulting syntheses, the equivalent width contribution functions of the three molecular bands used in the 3D analysis are also given in Fig.~\ref{fig:MARCS} (bottom three panels). The change in atmosphere and model chemistry has very little effect on the formation and strength of the carbon bands. As expected, the OH-band behaves differently as the change in $\oa$ directly changes its formation and strength. However, the differences between \sab\ and \smb\ and the differences between \sac\ and \smc\ are too small to change $\oa$ that would be determined by a comparative fit of them.

\begin{figure}[!h]
\begin{center}
\includegraphics[width=0.95\linewidth]{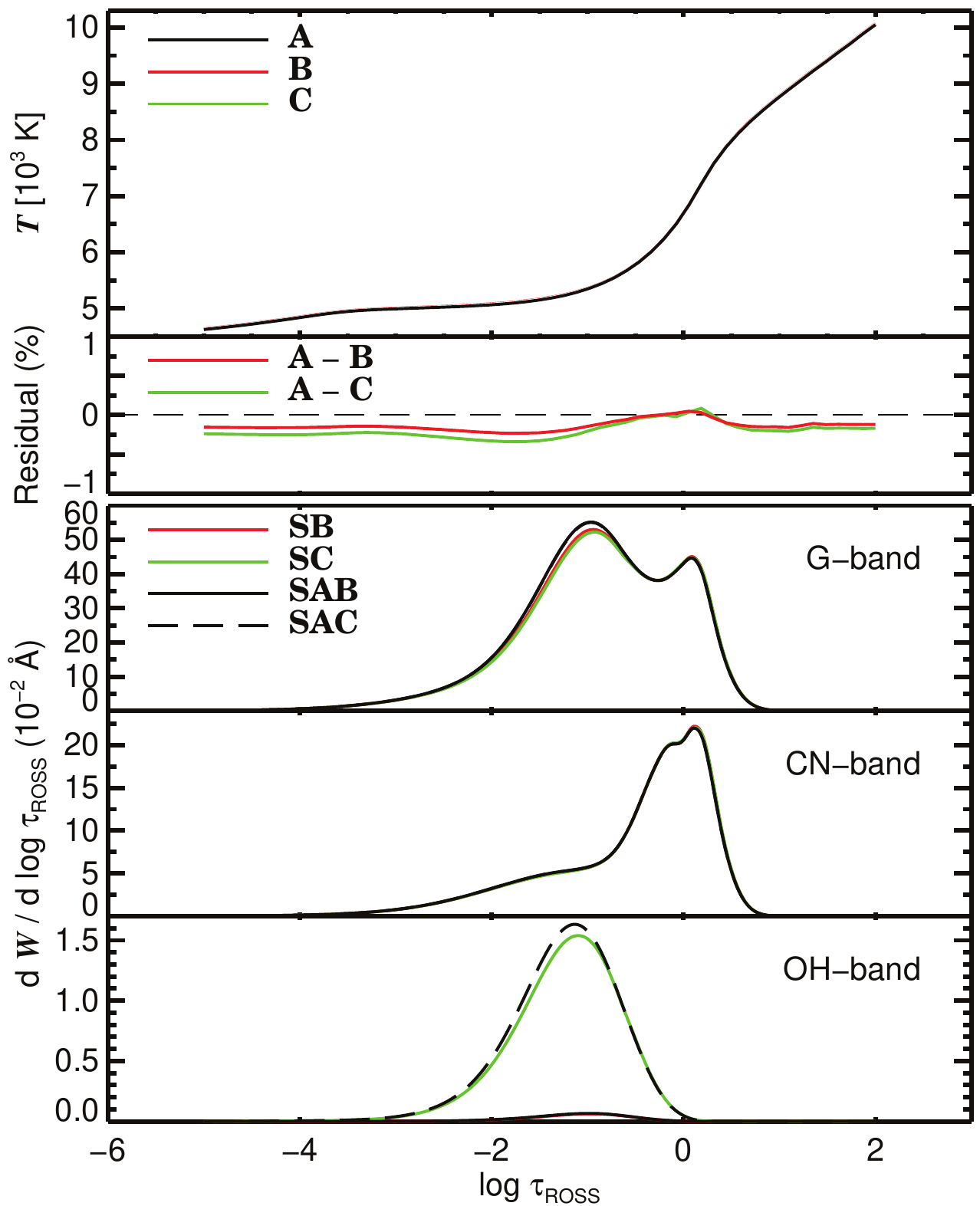}
\end{center}
\caption{1D ATLAS12 temperature structures for the model abundances given in Table~\ref{tab:models} and the residuals (top two panels). Resulting contribution functions for the CH- CN- and OH-bands are given in the bottom three panels. The contribution functions for the CN-band are indistinguishable.}
\label{fig:ATLAS}
\end{figure}

When we conduct the same tests with the ATLAS12 models (Fig.~\ref{fig:ATLAS}), we find extremely similar result. In fact, the differences in the temperature structures between \ma, \mb\ and \mc\ are even smaller than were found in Fig.~\ref{fig:MARCS} in the outer most layers of the models. The equivalent width contribution functions from the three molecular bands are also remarkably similar to those computed using the OSMARCS models and the differences between \sab\ and \smb, and between \sac\ and \smc\ are virtually indistinguishable.

The subtle differences found between the ATLAS12 and OSMARCS model structures are too small to significantly influence the resulting spectra computed from them as the largest deviations are found in regions in the model where line formation does not take place. Based on the data presented above, we conclude that the intrinsic chemical structure of a 1D model is unimportant to the resulting syntheses used to model CEMP star spectra, and that artificially enhancing the relevant CNO abundances during the spectral synthesis process is sufficient in the 1D caveat at this temperature regime. This is because CNO opacities have no strong effect on the temperature structure, making it independent of their abundances.

\end{appendix}

\end{document}